\documentclass[aps, pra, a4paper, amsfonts, amssymb, amsmath, reprint, showkeys, nofootinbib, twoside,superscriptaddress]{revtex4-2}
\usepackage[english]{babel}
\usepackage[utf8]{inputenc}
\usepackage{amsmath}
\usepackage{csquotes}
\usepackage{graphicx}
\usepackage[justification=raggedright, singlelinecheck=false,
            labelsep=period]{caption}

\AtBeginDocument{}
\usepackage{subcaption}
\usepackage{braket}
\usepackage{enumitem}
\usepackage{tikz}
\usepackage{comment}
\usepackage{tkz-berge}
\usepackage{xcolor}
\usepackage[mode=buildnew]{standalone}
\usepackage{float}
\usepackage{placeins}
\usepackage[colorinlistoftodos, color=green!40, prependcaption]{todonotes}
\usepackage[pdftex, pdftitle={Article}, pdfauthor={Author}]{hyperref} 
\usepackage{xcolor}
\definecolor{MyCustomColor}{RGB}{204, 0, 102}

\usepackage[colorinlistoftodos, color=green!40, prependcaption]{todonotes}
\usepackage{todonotes}

\begin{document}
\title{Enhanced quantum metrology with robust multipass interferometry}

\author{Sayak Mukherjee}
    \affiliation{Sussex Centre for Quantum Technologies, University of Sussex, Brighton, BN1 9QH, United Kingdom}
\author{José Afonso Oliveira}
    \affiliation{Sussex Centre for Quantum Technologies, University of Sussex, Brighton, BN1 9QH, United Kingdom}
\author{Sean William Moore}
    \email[Corresponding author email address: ]{sean.moore@lip6.fr} 
    \affiliation{Sussex Centre for Quantum Technologies, University of Sussex, Brighton, BN1 9QH, United Kingdom}
    \affiliation{Sorbonne Universit\'e, CNRS, LIP6, F-75005 Paris, France}
\author{Jacob A. Dunningham}
    \affiliation{Sussex Centre for Quantum Technologies, University of Sussex, Brighton, BN1 9QH, United Kingdom}

\date{September 1, 2026}

\begin{abstract}

\noindent Quantum metrology typically uses entangled states to achieve measurement precisions beyond the standard quantum limit. The advantage increases with the size of the entangled state, however generating and preserving large entangled states remains a major experimental challenge. Multipass protocols offer an alternative approach by allowing a single probe to interact repeatedly with the parameter of interest, but their performance is highly susceptible to loss, which accumulates over successive passes and rapidly erodes the quantum advantage. Here we introduce a hybrid strategy that combines small, loss-resilient entangled states with multipass interferometry. We show that this approach retains the robustness of small entangled probes while exploiting repeated interactions to achieve substantial enhancements in measurement precision. Furthermore, we propose a concrete implementation using currently available technologies, demonstrating that the predicted performance gains should be experimentally accessible with existing capabilities.

\end{abstract}


\maketitle

\section{Introduction}

Quantum metrology exploits uniquely quantum resources, including entanglement and squeezing, to achieve measurement precisions that surpass the limits attainable with classical systems~\cite{Giovannetti2006,Braunstein1994,Pezze2018}. As the field has matured from theoretical proposals to practical technologies, quantum-enhanced sensing has found applications in areas including gravitational-wave detection~\cite{LIGO2013}, atomic clocks and frequency standards~\cite{Ludlow2015}, electric and magnetic field sensing~\cite{Degen2017}, quantum imaging~\cite{Moreau2019}, and inertial sensing and navigation~\cite{Kasevich2002}. A central challenge is to retain these quantum advantages under realistic experimental conditions, where imperfections such as particle loss and decoherence can rapidly degrade performance.

Phase estimation provides the canonical setting for quantum metrology. In the absence of noise, maximally path-entangled NOON states~\cite{Dowling2008} are among the optimal probe states because they maximise the quantum Fisher information for a fixed particle number and achieve Heisenberg-limited precision, with an uncertainty that scales as $1/N$, compared with the standard quantum limit of $1/\sqrt{N}$ obtained using unentangled particles~\cite{Braunstein1994,Giovannetti2006}. In principle, this scaling favours ever-larger entangled states. However, in practice, generating high-fidelity optical NOON states remains extremely challenging, with experiments currently limited to relatively small particle numbers~\cite{Afek2010}. Consequently, the practical advantages of Heisenberg-limited probes are often difficult to realise.

An elegant alternative is provided by multipass interferometry, in which a single photon interacts repeatedly with the unknown phase so that the phase shift accumulates coherently over many passes~\cite{Higgins2007}. The number of passes then plays a role analogous to the number of particles in a NOON state, allowing Heisenberg-like scaling without requiring the preparation of large entangled states~\cite{Braun2018, Juffmann2016}. Multipass protocols are therefore considerably more accessible experimentally and have also been extended to distributed sensing \cite{Liu2021}. Unfortunately, they inherit the same fundamental vulnerability to loss. Since the photon must survive every traversal of the interferometer, the success probability decreases exponentially with the number of passes, rapidly eroding the metrological advantage in realistic systems~\cite{Demkowicz2012,Demkowicz2015,Escher2011}. Thus, despite their different physical implementations, large NOON states and single-photon multipass schemes ultimately suffer from the same limitation when loss is present.

A complementary approach has been to engineer probe states that sacrifice some ideal precision in exchange for substantially improved robustness against particle loss~\cite{Dorner2009,Kacprowicz2010, Knysh2011,Jarzyna2013}. Such states generally do not achieve the maximum quantum Fisher information in the absence of noise, but their greater resilience allows them to outperform NOON states once realistic levels of loss are taken into account. However, this strategy retains an important practical obstacle: increasing the achievable precision still requires the preparation of progressively larger entangled states, which remains experimentally demanding.

In this work, we combine the strengths of these two approaches. We consider small, experimentally accessible entangled states that are intrinsically more robust to loss than NOON states, and embed them within a multipass interferometric protocol so that repeated coherent interactions amplify the accumulated phase while maintaining the improved resilience of the probe. We show, through an analysis based on the quantum Fisher information, that this hybrid strategy can significantly outperform both conventional NOON-state interferometry and its single-photon multipass analogue under realistic loss. The framework is applicable to a broad class of loss-robust entangled states. To demonstrate experimental feasibility, we further analyse a concrete implementation based on four-photon entangled states together with a practical measurement scheme, showing that the predicted enhancement lies within the reach of current photonic technology.

\section{Multipass Mach-Zehnder interferometer} \label{sec:multipass}

\begin{figure}
    \centering
    \includegraphics[width=\linewidth, height = 5.5cm]{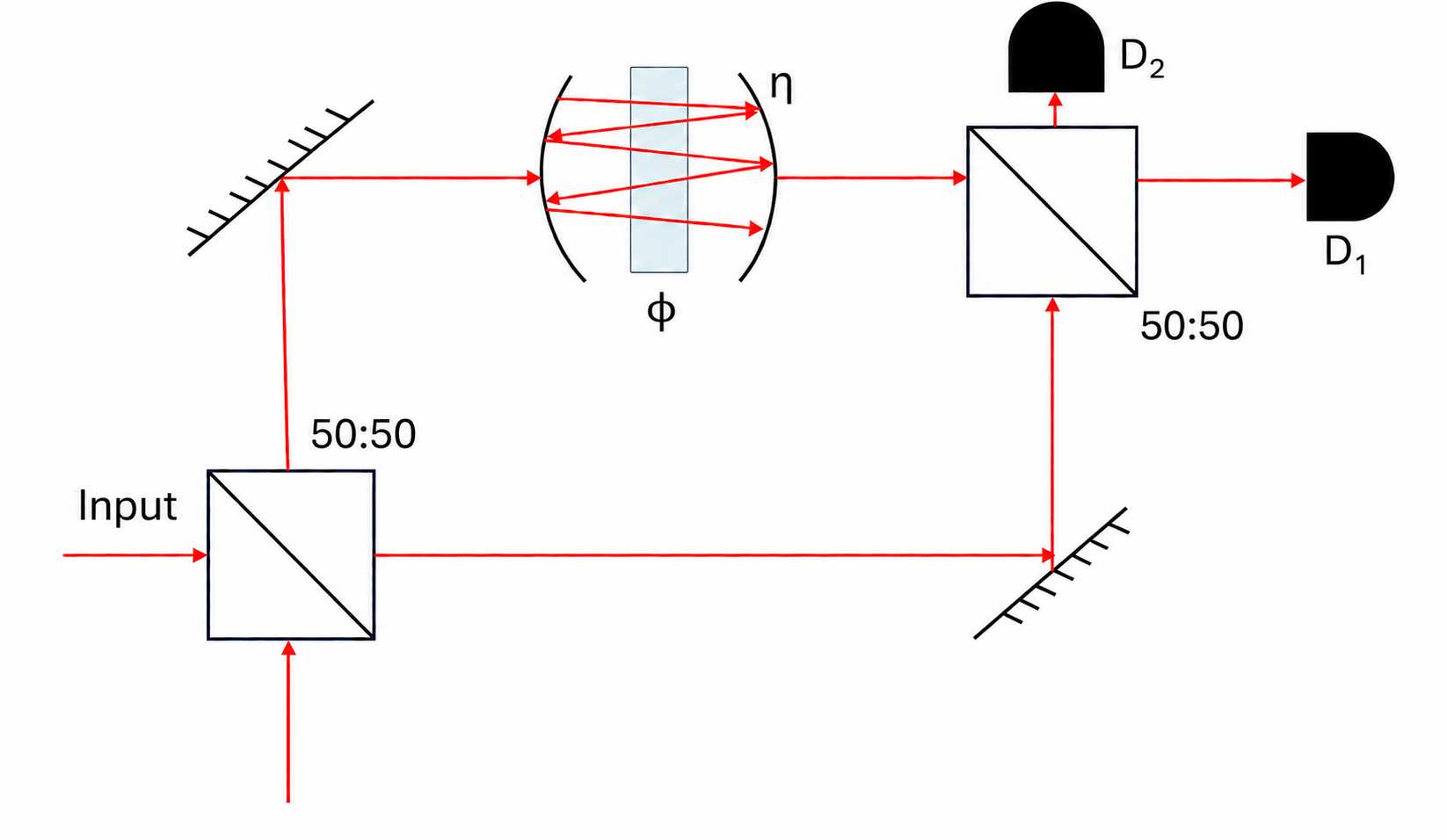}
    \caption{The multipass Mach-Zehnder interferometer consisting of two 50:50 beam splitters, two mirrors, photon number counting detectors $D_1$ and $D_2$ and an optical cavity. The upper path is passed multiple times through the sample in the cavity, accumulating a phase, $\phi$, each time. Environmental losses are  modelled by considering the cavity mirrors to be leaky with a reflection probability of $\eta$.}
    \label{fig:multipass MZI diagram}
\end{figure}

Our setup consists of a Mach-Zehnder interferometer incorporating a multipass element in one arm, as shown in Fig.~\ref{fig:multipass MZI diagram}. We assume that a sufficiently large number of independent measurements can be performed so that asymptotic estimation theory applies, rather than the finite-data regime considered in Refs.~\cite{Rubio2019,Rubio2020}. In this limit, the variance of any unbiased estimator is bounded by the quantum Cram\'er-Rao bound, (CRB),
\begin{equation}
    \delta \phi \geq \frac{1}{\sqrt{\mu F_{Q}}},
\end{equation}
where $\delta \phi$ is the phase uncertainty, $\mu$ is the number of measurements made (which is assumed to be large) and $F_{Q}$ is the quantum Fisher information (QFI). The QFI is the maximum Fisher information attainable over all possible quantum measurements (POVMs) performed on the probe state and therefore determines the ultimate precision permitted by quantum mechanics for that state. For single-parameter estimation, the quantum CRB is always saturable: there exists a measurement corresponding to projections onto the eigenbasis of the symmetric logarithmic derivative (SLD) that achieves the QFI locally around the true parameter value~\cite{Braunstein1994,Barndorff2000}. In practice, these optimal measurements may be difficult to implement and a practical (but suboptimal) measurement may need to be used instead. Making a particular choice of measurement yields the classical Fisher information, $F_c$, which is bounded from above by the QFI, i.e.
\begin{equation}
    F_C \leq F_{Q}.
    \label{eq:Fc_bound}
\end{equation}

Often in quantum metrology we are interested in finding measurements that come close to saturating the QFI while remaining experimentally accessible. We will come back to this point later in the paper but, for now, we focus on the QFI with the knowledge that the measurement required to saturate it does exist.

We begin by considering the NOON state, which is optimal for ideal lossless phase estimation with a fixed number of particles and therefore provides a natural benchmark against which to compare practical schemes. This state takes the form
\begin{equation}
    \ket{\psi} = \frac{1}{\sqrt{2}} \left( \text{e}^{iN \phi}\ket{N, 0} + \ket{0, N} \right),
    \label{NOON}
\end{equation}
where the $N$ atoms on the first path have each acquired a phase $\phi$, giving an overall phase of $N\phi$. 

The QFI for a general pure state $|\psi(\phi)\rangle$ is given by
\begin{equation}
F_{Q} = 4\left( \langle \psi '(\phi)| \psi '(\phi)\rangle - |\langle \psi '(\phi)| \psi (\phi)\rangle|^2 \right), \label{eq:fisher_pure}
\end{equation}
where $|\psi'(\phi)\rangle \equiv \partial|\psi(\phi)\rangle/\partial\phi$. From this, the QFI for the NOON state is $F_{Q}=N^2$ and so the precision of the phase given by the CRB is $\delta\phi \leq 1/(\mu N)$. This is known as the Heisenberg scaling. In principle, increasing $N$ continually improves the achievable precision~\cite{Berry2009,Gorecki2020}. In practice, however, preparing large optical NOON states with high fidelity is extremely challenging, and their extreme sensitivity to photon loss rapidly erodes their metrological advantage~\cite{Dowling2008}.

Higgins et al. proposed an elegant alternative in which a single photon repeatedly traverses the phase element, allowing the phase shift to accumulate coherently over many passes rather than being acquired simultaneously by many entangled photons~\cite{Higgins2007}. For $p$ passes, the resulting state just before the final beam splitter is,
\begin{equation}
    |\psi(\phi)\rangle = \frac{1}{\sqrt{2}}\left(e^{ip\phi}|1,0\rangle + |0,1\rangle\right),
\end{equation}
and, using Eq.~\ref{eq:fisher_pure}, the QFI can be shown to be $F_{Q}=p^2$. The resulting QFI is therefore identical to that of an $N$-photon NOON state after the identification of $p$ with $N$, i.e. the number of coherent passes plays precisely the role of the particle number. Consequently, Heisenberg scaling can be achieved without the need to generate large entangled states. 

Although it now takes multiple passes to encode a phase that the NOON state could do with a single pass, the practical advantages of dealing with a single particle rather than a large entangled state are substantial. Moreover, after the final beam splitter, conventional photon counting constitutes an optimal measurement for this state, so that the resulting classical Fisher information saturates the QFI.

Higgins et al. \cite{Higgins2007} demonstrated experimentally that repeated coherent interactions can replace multipartite entanglement as the resource responsible for Heisenberg-limited phase estimation~\cite{Higgins2007}. Their adaptive multipass protocol achieved the same $1/p$ precision scaling as an ideal $p$-photon NOON state while requiring only a single photon and linear optics. The work established multipass interferometry as a practical alternative to generating large entangled states, although the requirement that the photon survive every traversal leaves the protocol highly susceptible to loss.

\begin{figure}[b]
    \centering
    \includegraphics[width=\linewidth]{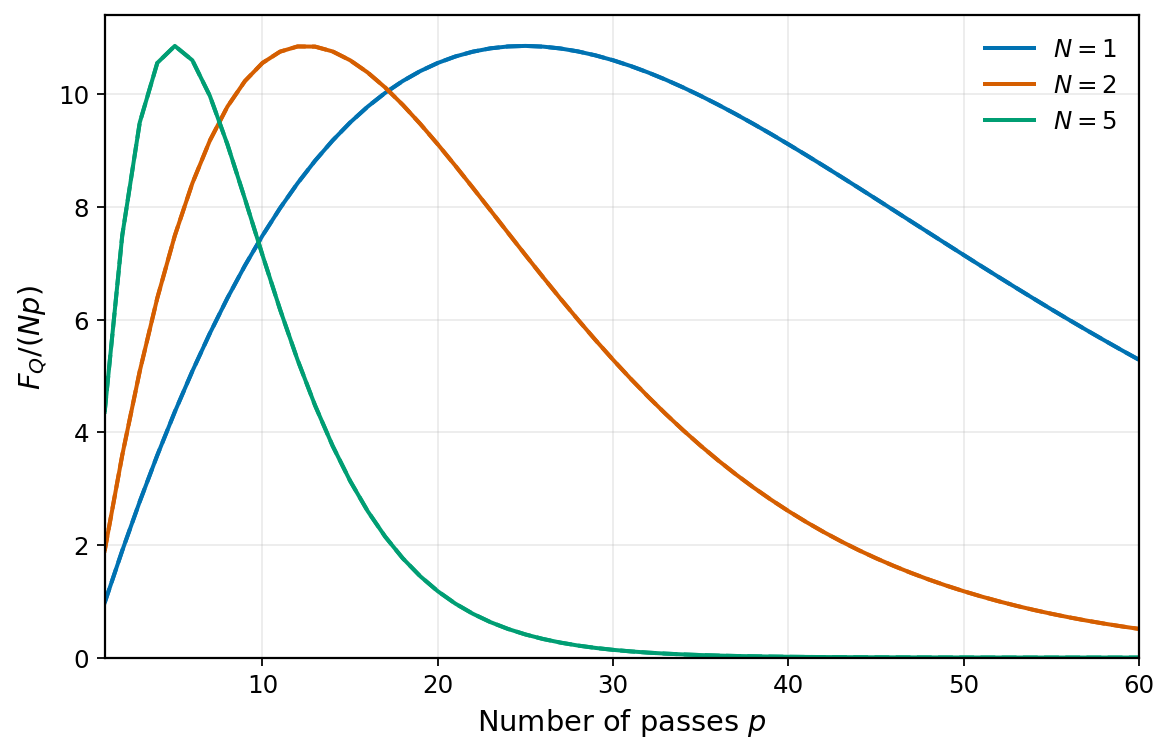}
    \caption{The quantum Fisher information per particle-pass, $F_Q/(Np)$, is shown as a function of the number of passes, $p$, for NOON states with $N=1,2,5$. In each case $\eta=0.95$. All the different sized NOON states achieve the same maximum value.}
    \label{fig:NOON_comparison}
\end{figure}

\section{Lossy Regime}

Particle losses are typically modelled by using a fictitious beam splitter to skim off some particles into an environmental mode which is then traced over. In our scheme, we can model loss by taking the mirrors in the cavity (see Figure \ref{fig:multipass MZI diagram}) to have imperfect reflectivity $\eta<1$. We assume that loss occurs only in the sensing arm containing the sample. This reflects the common situation in which attenuation arises from interaction with the sample, while the reference arm can be engineered to be effectively lossless \cite{Dorner2009,Demkowicz2015}.

If we take the creation operator for a particle on the upper path to be $\hat{a}^{\dagger}$, and for the lower (reference) path to be $\hat{b}^{\dagger}$, the loss transformations are
\begin{eqnarray}
    &&\hat{a}^{\dagger} \xrightarrow{} \sqrt{\eta} \: \hat{a}^{\dagger} + \sqrt{1 - \eta} \: \hat{c}^{\dagger} \\
    &&\hat{b}^{\dagger} \xrightarrow{} \hat{b}^{\dagger},
\end{eqnarray}   
where $c^\dagger$ corresponds to the environmental mode that is discarded.

A similar scenario was investigated in \cite{Demkowicz2009} but without the multipass cavity. They considered a general $N$-particle two-mode state of the form
\begin{equation}
  \ket{\psi} = \sum_{k=0}^{N} \alpha_k \ket{k, N - k},  
  \label{eq: generic state}
\end{equation}
where $\ket{k, N - k}$ represents $k$ particles on the upper path of the interferometer and $N-k$ particles on the lower path. They showed that particle loss commutes with phase imprinting and so it makes no difference if the particles are lost before, during, or after the phase acquisition. The QFI was found to be
\begin{equation}
  F_Q= 4\left[ \sum_{k=0}^{N} k^2 |\alpha_k|^2 - \sum_{l=0}^{N}\frac{\left(\sum_{k=l}^{N}|\alpha_k|^2\, k\, B_{l0}^{k}\right)^2}{\sum_{k=l}^{N}|\alpha_k|^2\, B_{l0}^{k}} \right], 
  \label{eq:lossy QFI analytical}
\end{equation}
where
\begin{equation}
    B_{l0}^k = \left(\begin{array}{c} k \\ l \end{array} \right) \eta^{k-l} (1-\eta)^l.
\end{equation}

This result can be extended to the multipass setting by noting that there will be loss on each pass and so the fraction that is not lost transforms as $\eta \xrightarrow{} \eta^{p}$. The phase will also coherently add up on each pass meaning that the overall expression for $F_Q$ gains a factor of $p^2$.

\section{Results for Different States}

This expression can be solved analytically for NOON states to give,
\begin{equation}
F_{Q} = 2N^2 p^2\frac{\eta^{Np}}{1 + \eta^{Np}}.
\end{equation}
Here we see the equivalence of the number of particles, $N$, and the number of passes, $p$. Since they always appear as the product $Np$ we can trade one off against the other.

The figure of merit we are most interested in is the QFI scaled by the particle flux, $F_Q/(Np)$, because in practice we want to extract the maximum metrological information from a system for a given number of queries \cite{Zwierz2010}. The number of queries in our case is the number of particle-passes, i.e. $Np$. Such an approach is especially  important in quantum metrology schemes involving delicate samples \cite{Taylor2013} where we want to maximise the measurement precision while minimising the damaging flux through the sample.

In Fig.~\ref{fig:NOON_comparison} $F_Q/(Np)$ is shown as a function of $p$ for different sized NOON states, $N=1,2,5$. In each case $\eta = 0.95$. We see that $F_Q/(Np)$ reaches a maximum before tailing off. This is what we would expect because there is competition between the phase wind-up which increases the QFI for more passes, and the loss which degrades the QFI with more passes. For small NOON states more passes are needed to reach the maximum value, but these states are easier to make. Ultimately, all the different sized NOON states reach the same maximum value of $F_Q/(Np)$ so, for metrological purposes, are equivalently good.

\begin{figure}
    \centering
    \includegraphics[width=\linewidth]{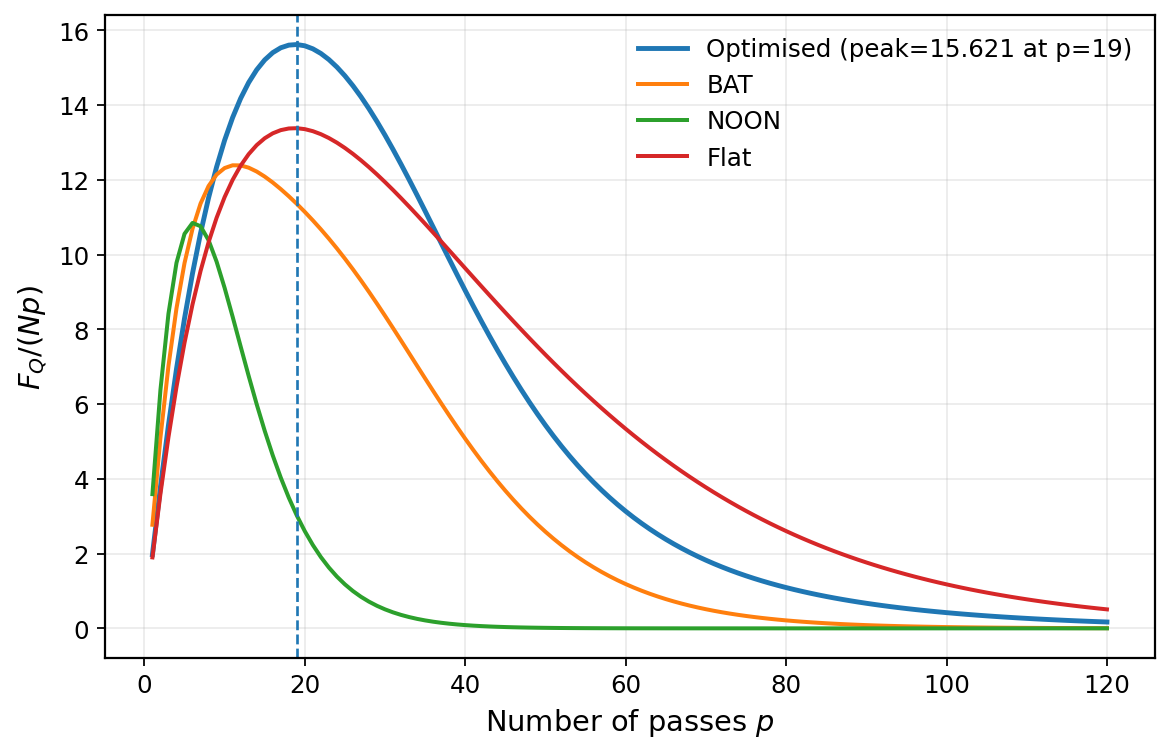}
    \caption{$F_Q/(Np)$ plotted as a function of $p$ for different states. The general state (\ref{eq: generic state}) with optimised coefficients is shown as the blue line. This is compared with the other states discussed in the text. In each case $N=4$ and $\eta=0.95$.}
    \label{fig:best theoretical QFI}
\end{figure}

The question we want to address is whether it is possible to surpass this limit by using other more-robust entangled states. We will consider the particular examples of the dual-Fock (or BAT) state \cite{Holland1993,Dunningham2002} and the flat state. After traversing the multipass cavity, these states have the respective forms,
\begin{equation}
\ket{\psi_{\mathrm{BAT}}}
= \sum_{n=0}^{N/2}
A_n e^{i 2n p \phi}
\ket{2n, N-2n},
\end{equation}
where 
\begin{equation*}
A_n
=
\frac{(-1)^n}{2^{N/2}}
\frac{\sqrt{(2n)!(N-2n)!}}
{n!\left(\frac{N}{2}-n\right)!}.
\end{equation*}
and 
\begin{equation}
\ket{\psi_{\mathrm{FLAT}}}
=
\frac{1}{\sqrt{N+1}}
\sum_{k=0}^{N}
e^{i k p \phi}
\ket{k,N-k}.
\end{equation}.

In order to keep the scheme practical, we restrict ourselves to small entangled states  with $N=4$, though further advantages will follow in the future when larger states can be generated. Four-photon BAT \cite{Sun2008} and NOON \cite{Nagata2007} states have already been realised experimentally and used for quantum-enhanced phase metrology. To our knowledge, flat states have not yet been experimentally realised, but they serve as a useful theoretical comparison here.

The results for these states are shown in Fig.~\ref{fig:best theoretical QFI}. We see that the flat and BAT states outperform the NOON state, achieving higher maximum values of $F_Q/(Np)$. This suggests a route to improved quantum-enhanced measurements in lossy environments. For comparison, the best possible state of the form of Eq.~(\ref{eq: generic state}) with $N=4$ is also plotted, where the values of $\alpha_k$ have been optimised to give the largest possible peak value of $F_Q/(Np)$. The optimised state serves as a useful benchmark here but, in general, will not be practical to make, so we will neglect it in the remainder of the paper.

An alternative approach is to modify NOON states to make them more resilient to losses. One way to do this is to pass a NOON state through a beam splitter with variable reflectivity that can be tuned to optimise the maximum value of $F_Q/(Np)$, which will depend on $\eta$. This means that we are optimising over both the number of passes, $p$ and the beam splitter reflectivity. We call the resulting state a rotated NOON state. It has the general form,
\begin{equation}
    \ket{\psi} = \frac{1}{\sqrt{2}} \sum_{k=0}^{N}\, c_k\,\left( \ket{N - k, k} + \ket{k, N - k}\right).
    \label{eq:modified_NOON}
\end{equation}

The intuition behind how this might help is that, for $c_k$ peaked near $k=0$, (\ref{eq:modified_NOON}) is still close to a NOON state and so will retain a large QFI and high measurement precision, but now the shuffling of photons between the paths means that the loss of a photon from a particular path no longer unambiguously collapses the state onto one of the terms in (\ref{eq:modified_NOON}) so does not destroy all the entanglement. This should give a good compromise between precision and robustness.

\begin{figure}[t]
    \centering
    \includegraphics[width=\linewidth]{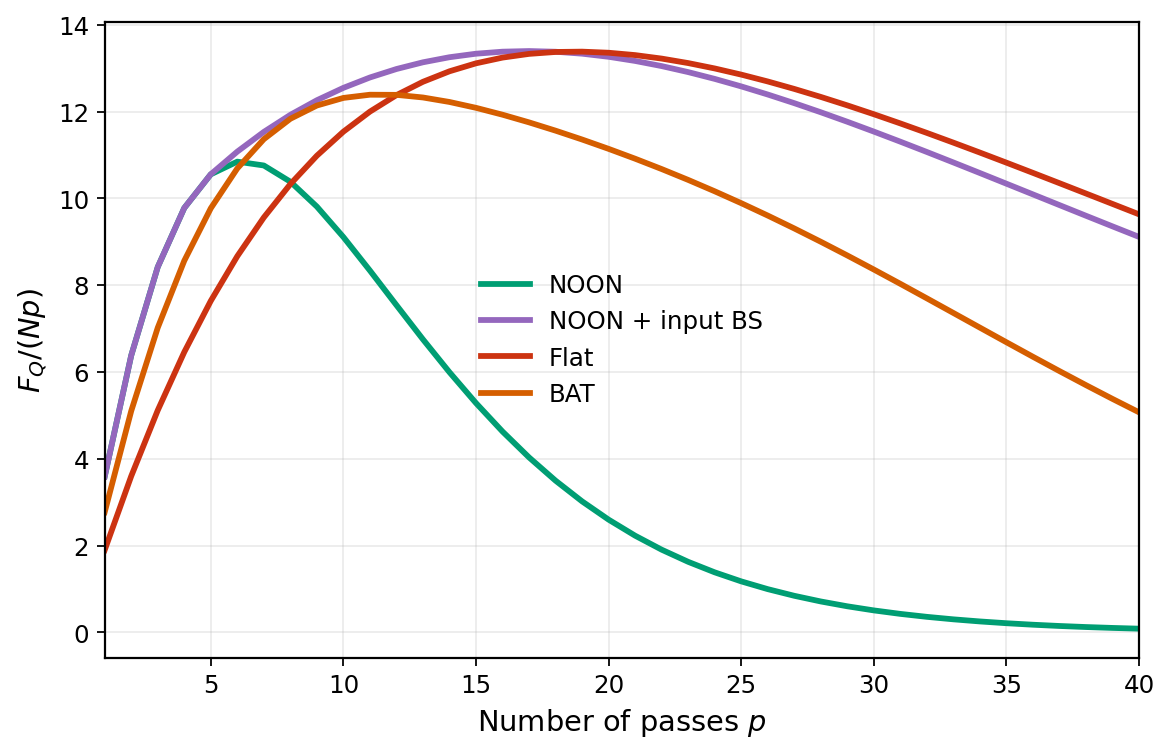}
    \caption{The results of Fig.~\ref{fig:best theoretical QFI} -- redisplayed with a different $p$ range and the optimised state removed -- are compared with the rotated NOON state. The rotated NOON state outperforms all the other states. For the parameters used here, $N=4$ and $\eta=0.95$, the optimal number of passes is $p = 17$ and the optimal beam splitter reflectivity is $0.83$.}
    \label{fig:QFI optimised plot}
\end{figure}

The results for the rotated NOON state -- optimised over the beam splitter's reflectivity -- are shown in Fig.~\ref{fig:QFI optimised plot} and compared with the results shown in Fig.~\ref{fig:best theoretical QFI}. For clarity we have used a different range of $p$-values and have removed the optimised state. The rotated NOON state clearly outperforms the NOON and BAT states and marginally outperforms the flat state. It also has the advantage that, unlike the flat state, it can be created in the laboratory with current technology.

\section{Practical Measurements}

The results so far show that, in principle, we can obtain enhanced measurement precision by combining small robust entangled states with a multipass interferometer. However, these are all in terms of the QFI. In practice we need to make a particular read-out measurement The classical distribution of data from the read-out measurements gives the classical Fisher information, $F_C$, which is bounded from above by $F_Q$ (see Eq.~\ref{eq:Fc_bound}). While it is known that for a single parameter, $\phi$, there exists a measurement that saturates this bound, the POVMs required to do so may be quite impractical.

\begin{figure}[h]
    \centering
    \includegraphics[width=\linewidth]{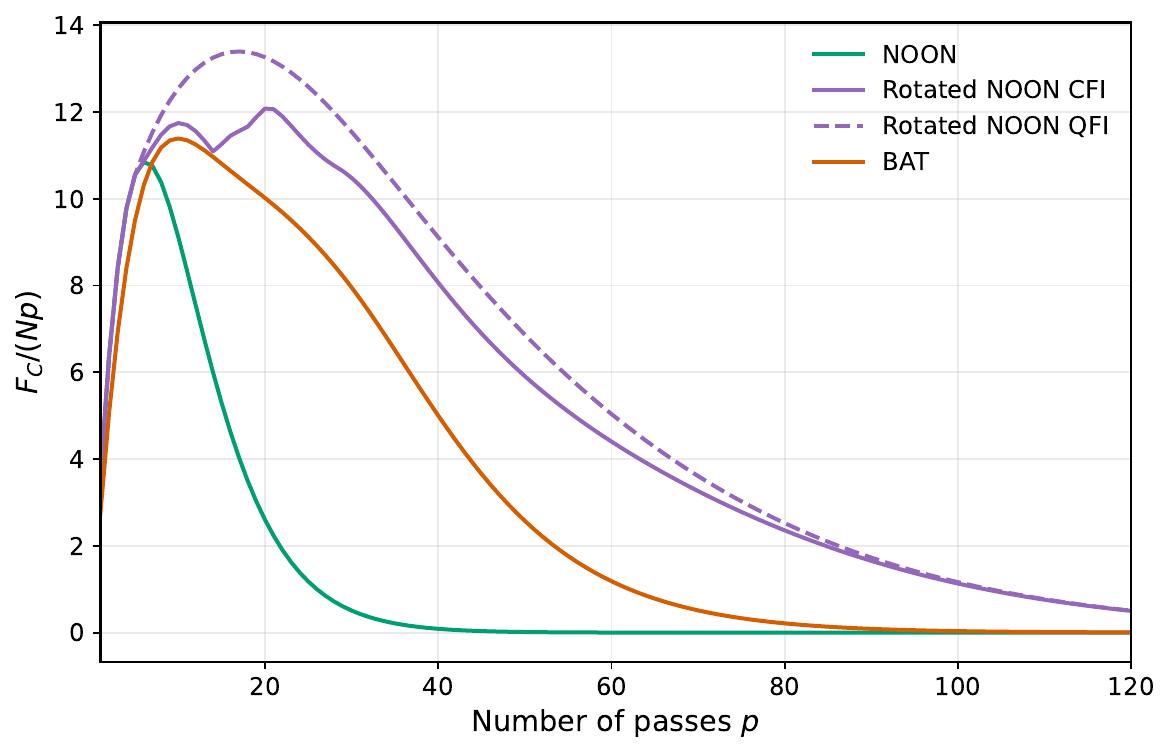}
    \caption{$F_C/(Np)$ is plotted as a function of, $p$, for different states. The flat state has been removed as the focus is on practical schemes that can be realised. $F_c$ is the classical Fisher information corresponding to the read-out measurement detailed in the text, i.e. a beam splitter with optimised reflectivity followed by number counting. The rotated NOON state, outperforms both the BAT and standard NOON states. The dashed line is the QFI for the optimal rotated NOON state for comparison.}
    \label{fig:optimal CFI extracted}
\end{figure}

Here we take a pragmatic approach and consider an experimentally-accessible read-out to see how well this performs. In particular, we consider passing the states through a beam splitter and counting particle numbers at the outputs as shown in Fig.~\ref{fig:multipass MZI diagram}. For a 50:50 beam splitter this is already an optimal measurement for the simple NOON states, but not for the BAT, flat, or rotated NOON. To improve things, we allow the beam splitter to have adjustable reflectivity. The performance of this scheme also depends on the local operating point $\phi$. Our problem, therefore, becomes an optimisation problem over $\phi$, the reflectivity of the output beam splitter, and (for the rotated NOON state) the reflectivity of the input beam splitter. This optimisation is repeated for each value of $p$.

\begin{figure}[h]
    \centering
    \includegraphics[width=\linewidth]{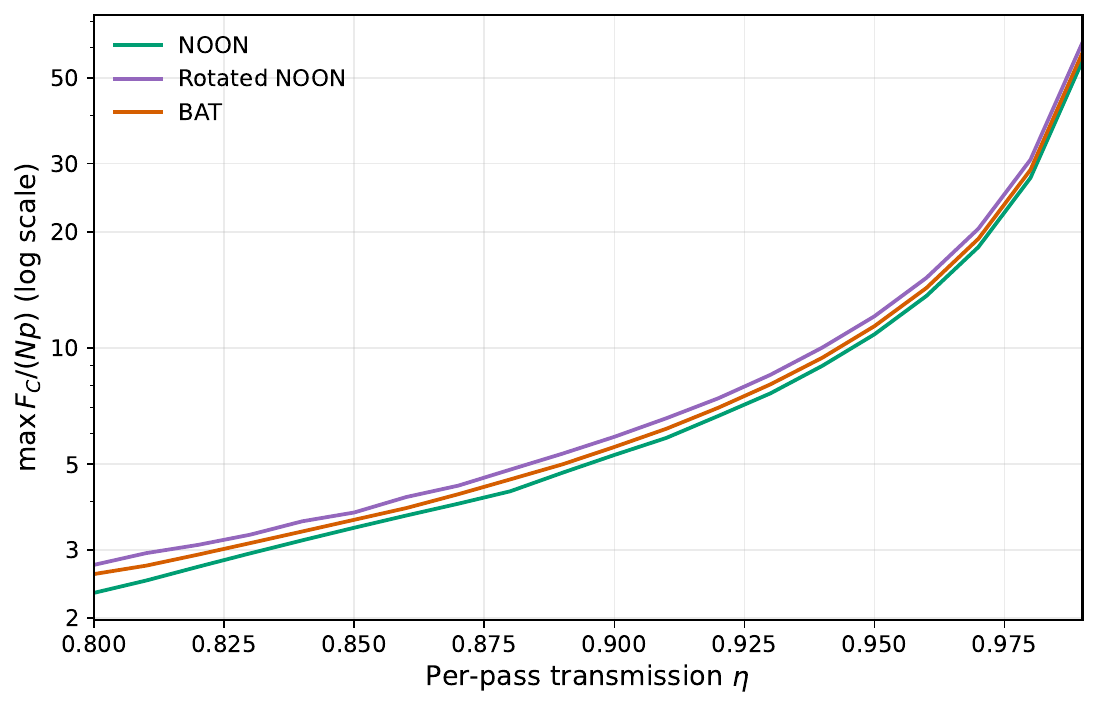}
    \caption{$\max_{p} F_C/(Np)$ is plotted as a function of $\eta$, the per-pass transmission. There is a clear hierarchy where the rotated NOON dominates the BAT which in turn dominates the NOON for all $\eta$ displayed.}
    \label{fig:optimal CFI over eta}
\end{figure}

The resulting classical Fisher information is scaled by the number of particle-passes and plotted as a function of $p$ for $\eta =0.95$ in Fig.~\ref{fig:optimal CFI extracted} and as a function of $\eta$ maximised over all $p$ in Fig.~\ref{fig:optimal CFI over eta}. We see that both the BAT and rotated NOON states surpass the simple NOON case. Both scenarios are experimentally feasible with current technology, but the rotated NOON is more appealing in terms of the advantage it offers.

\section{Conclusion}
We have shown that combining small, loss-robust entangled states with multipass interferometry provides a practical route to quantum-enhanced metrology in realistic lossy environments. The hybrid strategy outperforms conventional NOON-state schemes when the available resource is the number of particle-passes, while remaining compatible with experimentally accessible four-photon states and simple photon-counting measurements. An experimental implementation would need to be able to control the number of passes, $p$, but this could employ techniques developed in multipass microscopy \cite{Juffmann2016}. Although we have intentionally focused on a proof-of-principle implementation using small entangled states and non-optimised read-outs, the framework is general and leaves considerable scope for further improvements through optimised probe states, improved measurement strategies, and larger entangled resources. These results suggest that robust multipass interferometry provides a promising new direction for practical quantum sensing beyond the standard NOON-state paradigm.

\section*{Acknowledgements}
The authors acknowledge funding for this project from the United Kingdom EPSRC through the Quantum Information Science and Technology Centre for Doctoral Training. The authors thank Jackson Phoong for helpful comments on the manuscript.

\end{document}